%% file: pedersen-ec.tex
\documentclass[a4paper]{llncs}

\usepackage[backgroundcolor=black!10,bordercolor=white!70!black,linecolor=white!70!black]{todonotes}
\input{easycrypt-common}
\usepackage{breakcites}
\usepackage{bold-extra}
\usepackage{varioref}
\usepackage{tabularx}
\usepackage{textgreek}
\usepackage{amsfonts}
\usepackage{mathtools}
\usepackage{array}
\usepackage{xhfill}
\usepackage{arydshln}
\usepackage{listings}
\usepackage[lined]{algorithm2e} 
\usepackage{semantic}

\newcommand{\insertauthor}{Roberto Metere}
\newcommand{\inserttitle}{Automated Cryptographic Analysis of the Pedersen Commitment Scheme}

\PassOptionsToPackage{hyphens}{url}\usepackage{hyperref} 
\hypersetup{
  pdfauthor={\insertauthor},%
  pdfkeywords={formal methods,cryptography,secure computation,commitment,easycrypt},%
  pdftitle={\inserttitle},%
  colorlinks,
  linkcolor={red!50!black},
  citecolor={blue!50!black},
  urlcolor={blue!50!green!90!black},
  bookmarksopen,
  bookmarksdepth=2
}

\newlength{\protocolArrowLength}
\newcolumntype{B}[1]{>{\centering\arraybackslash}b{#1}}

\newcommand{\inr}{\in_{\mathrm{R}}}
\newcommand{\pbrackets}[1]{\left ( #1 \right )}
\newcommand{\sbrackets}[1]{\left [ #1 \right ]}

\newcommand{\set}[1]{\left \{ #1 \right \}}

\newcommand{\intset}{\mathbb{Z}}
\newcommand{\realset}{\mathbb{R}}
\newcommand{\groupset}{\mathbb{G}}
\newcommand{\boolT}{{\tt T}}

\newcommand{\zoset}{\set{0, 1}}

\newcommand{\pair}[2]{\left ( #1, #2 \right )}
\newcommand{\of}[1]{\left ( #1 \right )}

\newcommand{\alg}[1]{\mathcal{#1}}
\newcommand{\prob}[1]{\mathrm{Pr}\left[ #1 \right]}
\newcommand{\negligible}[1][n]{\mu\of{#1}}

\newlength{\reducingspace}
\newcommand{\reducespace}[1][\reducingspace]{\vspace{-#1}}
\newcommand{\reducespacetiny}{\reducespace[0.2\reducingspace]}
\newcommand{\reducespacesmaller}{\reducespace[0.4\reducingspace]}
\newcommand{\reducespacesmall}{\reducespace[0.7\reducingspace]}
\newcommand{\reducespacebig}{\reducespace[1.5\reducingspace]}

\begin{document}

\title{\inserttitle}

\titlerunning{\inserttitle}  
%
\author{Roberto~Metere, Changyu~Dong}
\authorrunning{Roberto Metere et al.} 
%
\tocauthor{Roberto Metere, Changyu Dong}
\institute{Newcastle University (UK)\\
\email{r.metere2@ncl.ac.uk}, \email{changyu.dong@ncl.ac.uk}%
}

\maketitle              

\input{sections/abstract}

\input{sections/intro}
\input{sections/relatedworks}
\input{sections/background}
\input{sections/contribution}
\input{sections/conclusion}

%
%
\bibliographystyle{splncs03}
\bibliography{bibliography}{}
%
\end{document}

%% file: easycrypt-common.tex
\usepackage{amsmath}

\newenvironment{tightcenter}{%
  \setlength\topsep{0pt}
  \setlength\parskip{0pt}
  \begin{center}}
{\end{center}}

\usepackage{xspace}

\usepackage{mathpartir}

\usepackage{textcomp}
\usepackage{listings}

\input{easycrypt.lst}

\lstnewenvironment{easycrypt}[3][]%
  {\lstset{language=easycrypt,caption={#2},label={#3},#1}}%
  {}

\lstnewenvironment{easycryptmath}[3][]%
  {\lstset{language=easycrypt-math,caption={#2},label={#3},#1}}%
  {}

\lstnewenvironment{easycryptfloat}[3][]%
  {\lstset{language=easycrypt,float=tp,caption={#2},label={#3},#1}}%
  {}

\newcommand{\ecinput}[5][]%
{\lstinputlisting[language=easycrypt,linerange={#4},caption={#3},label={#5},#1]{#2}}

\newcommand{\ecinputfloat}[4]%
{\lstinputlisting[language=easycrypt,float=tp,linerange={#3},caption={#2},label={#4}]{#1}}

\def\ec#1{\lstinline[language=easycrypt-math]"#1"}



%% file: sections/abstract.tex

\begin{abstract}
Aiming for strong security assurance, recently there has been an increasing interest in formal verification of cryptographic constructions.
This paper presents a mechanised formal verification of the popular Pedersen commitment protocol, proving its security properties of correctness, perfect hiding, and computational binding.
To formally verify the protocol, we extended the theory of EasyCrypt, a framework which allows for reasoning in the computational model, to support the discrete logarithm and an abstraction of commitment protocols.
Commitments are building blocks of many cryptographic constructions, for example, verifiable secret sharing, zero-knowledge proofs, and e-voting.
Our work paves the way for the verification of those more complex constructions.

\keywords{formal verification, cryptography, commitment, easycrypt}
\end{abstract}

%% file: sections/intro.tex
\section{Introduction}
\label{sec:intro}
\reducespacetiny

The high and increasing volume of communication exchanged through insecure channels, e.g. the Internet, confers increasing significance on security guarantees of cryptographic protocols.
However, the increasing complexity in the design of cryptographic protocols leads to more complex and long proofs, making them error-prone and difficult to check.
In fact, it happened that protocols believed to be secure, even for years, were found to be vulnerable to attacks after further investigation~\cite{lowe1995attack,abadi1997explicit,bouillaguet2011automatic,schmidt2012automated}.
As a result, it is highly desirable to have tools that can formally and automatically verify cryptographic protocols.
Recently, many tools have been developed based on different approaches for this purpose~\cite{meadows1996nrl,lowe1996breaking,goubault2000method,abadi2005computer,barthe2013verified,cremers2016automated}.
They have been proven to be effective in verifying security properties and finding attacks.

Most of the automatic tools aiming to aid proofs for cryptographic protocols~\cite{kemmerer1989analyzing,mitchell1997automated,blanchet2001efficient,ryan2001modelling,song2001athena,corin2002improved,armando2004satmc,turuani2006cl,escobar2007diffie,ramsdell2009cpsa,meier2010strong,armando2012avantssar,schmidt2012automated} work in the symbolic model, some providing computational soundness for special cases~\cite{backes2003composable}, while only few support the computational model~\cite{blanchet2008computationally,barthe2009formal,barthe2011computer,ambrona2016automated}.
The symbolic model~\cite{dolev1983security} describes the real world cryptographic protocols abstractly.
Messages are literals, therefore it is difficult to model partial leakage.
Cryptographic primitives are assumed to be perfect and used as black boxes, which is not realistic.
The adversaries strategies are often predefined and limited to the inference rules provided, therefore the proofs are limited to case-by-case reasoning~\cite{bana2014computationally}.
This overly simplistic way can capture errors in the logic of the design, but cannot fully describe situations in which the cryptographic primitives cannot be treated as black boxes and when the security properties are defined computationally, which is often true in cryptographic protocols and other fields of cryptography.
On the contrary, the computational model is more realistic when modelling cryptographic protocols and can capture many low level details which are needed in proofs.
Cryptographic primitives are secure in the sense that computationally bounded adversaries can break the security properties with only a negligible probability, if certain assumptions hold.
Messages are bit strings and adversaries can be any polynomial time probabilistic Turing machine.
Therefore, it allows for reasoning with probabilities and cryptographic assumptions.
This allows for rigorous proofs that cannot be obtained in the symbolic model.

In this paper we focus on proving properties of commitment schemes.
Commitment schemes are a cryptographic primitive that have been widely used by its own or as a building block in other protocols, for example, verifiable secret sharing, zero-knowledge proofs, and e-voting~\cite{naor1991bit,goldreich1996composition,schoenmakers1999simple}.
The high level concept behind commitment protocol is intuitively simple, a committer wants to commit to a message, while keeping it secret from a receiver until a later time when the committer reveals the message to the receiver.
In this paper we prove the security properties of the popular Pedersen commitment scheme~\cite{pedersen1991non}.
To the best of our knowledge, this is the first mechanised formalisation of this protocol.

Despite being a basic primitive in secure computation, to formalise it and have a computer generated proof is far from trivial.
In the security proof generated by humans, many small gaps are left by the prover as they are easy to prove.
However, for a machine the gaps can be huge and extra efforts need to be spent to let the machine complete the proof.
In particular, to prove the perfect hiding property, we created a sequence of games that vary slightly to allow the machine to carry on the proof.
This additional construction is totally absent from proofs in the original paper and textbooks.
In addition, to prove computational binding, we constructed a discrete logarithm game to allow for reduction.

%

The paper is organised as the following: in Section~\ref{sec:related_work} we review the related work.
In Section~\ref{sec:background}, we introduce the preliminaries.
We then describe the formal proofs in relation to the constructions, in Section~\ref{sec:contribution}.
In Section~\ref{sec:conclusion} we conclude and discuss possible future work.

%% file: sections/relatedworks.tex

\section{Related work}
\label{sec:related_work}
\reducespacetiny

To avoid the limitations of symbolic model tools and capture all the properties of the cryptographic primitive, we used a tool allowing for reasoning in the computational model.
Some frameworks are available to work in such a model.
CryptoVerif~\cite{blanchet2008computationally} is based on concurrent probabilistic process calculus.
Although it is highly automatic,
it is limited to prove properties related to secrecy and authenticity.
The tool gga\textsuperscript{$\infty$}~\cite{ambrona2016automated} specialises in reasoning in the generic group model and seems promising when attackers have access to random oracles, which does not apply to our setting.
Certicrypt is a fully machine-checked language-based framework built on top of the Coq proof assistant~\cite{barthe2009formal}.
However it is no longer maintained.
EasyCrypt~\cite{barthe2011computer} follows the same approach as CertiCrypt and supports automated proofs as well as interactive proofs that allow for interleaving both program verification and formalisation of mathematical theories.
This is desirable because they are intimately intertwined when formalising cryptographic proofs, and can leave the tedious parts of proofs to machines.


Recently, a machine-checked formalisation of \textSigma-protocols to prove statements about discrete logarithms  has been developed in  CertiCrypt~\cite{barthe2010machine}.
A commonality between the work in~\cite{barthe2010machine} and our work is that the Schnorr protocol proved in~\cite{barthe2010machine} is also based on the discrete logarithm assumption.




%% file: sections/background.tex

\section{Background}
\label{sec:background}
\reducespacetiny

\subsection{Commitment Protocols}
\label{sec:commitments_description}
\reducespacetiny
Commitment protocols are two-party schemes between a committer $C$ and a receiver $R$, and run in two phases.
Let $M$ be the space of messages to commit to.
The first phase is called {\em commitment phase}, where the party $C$ sends $R$ its commitment for a private message $m \in M$ and secretly holds an opening value.
The second phase is called {\em verification phase}, where the party $C$ sends $R$ the original message $m$ along with the opening value, so that $R$ can verify that the message committed in the first phase was indeed $m$.

Commitment protocols involve three efficient algorithms: (i) $\alg{G}\of{1^n}$ which outputs a public value $h$, (ii) $\alg{C}\of{h, m}$ which takes as input the public value $h$ and the message $m$, and outputs $\pair{c}{d}$ where $c$ is the commitment to send in the first phase and $d$ is the opening value to be send in the second phase, and (iii) $\alg{V}\of{h, m, c, d}$ which takes as input the public value $h$, the message $m$, the commitment $c$ and the opening value $d$, and outputs {\em true} if verification succeeds or {\em false} otherwise.
Let $\pi = (\mathcal{G}, \mathcal{C}, \mathcal{V})$ be a commitment scheme, its security properties are
(i) correctness, i.e. for every message the commitment generated is valid, (ii) computational or perfect hiding, where any attacker cannot learn information from the commitment $c$ about the message $m$ with any advantage (perfect), or with a negligible advantage (computational), and (iii) computational or perfect binding, where the message $m$ is uniquely bound to $c$ (perfect) or finding another message with the same commitment has negligible probability of success (computational).
We defer the formal definitions of these properties to Section~\ref{sec:contribution}.

\subsection{Reasoning in EasyCrypt}
\label{sec:easycrypt}
\reducespacetiny

We present a brief overview of EasyCrypt.
More information about Easycrypt and the syntax of its language can be found in~\cite{barthe2014easycrypt,easycryptmanual}.
Those who are familiar with EasyCrypt can skip the rest of this section.

EasyCrypt handles the computational model, in which adversaries are probabilistic algorithms.
To capture this, we have modules that are containers of global variables and procedures.
Procedures capture the idea of algorithm, and one can reason about procedures running in a memory (as an execution environment).
In the computational model, one has to reason about the probability of adversaries returning some specific results.
EasyCrypt captures this idea as the probability of running a procedure $M.c$ in a memory $m$ with post-condition $Q$ evaluating true, where $M$ is the module containing the procedure $c$, written as
\begin{center}
  \ec{Pr[$M.c$($\dots$) @ &$m$ : $Q$]}
\end{center}
To express and to prove properties, in EasyCrypt we use judgements (assertions) in (i) basic higher-order logic for implemented theories, (ii) Hoare logic (HL), (iii) probabilistic Hoare logic (pHL), and (iv) probabilistic relational Hoare logic (pRHL).
For the last three of them, there are concepts of pre-condition $P$ and post-condition $Q$, as well as procedures $M.c$, $M.c_1$, $N.c_2$, \dots, inside modules $M$, $N$, running in some memory $m$.
\begin{description}
  \item[HL]   \ec{hoare[$M.c$ : $P \Rightarrow Q$]}              - When $P$ is true relating to some memory $m$ and $M.c$ terminates in $m$, then after running $M.c$, $Q$ always evaluates to true in the (modified) memory.
  \item[pHL]  \ec{phoare [$M.c$ : $P \Rightarrow Q$] <\ $r$}     - When $P$ is true relating to some memory $m$ and $M.c$ terminates in $m$, then after running $M.c$, $Q$ evaluates to true in the (modified) memory with probability less than $r \in \sbrackets{0, 1} \subset \realset$.
  Other supported relations are the common relations $=$, $>$, $\geq$, and $\leq$.
  \item[pRHL] \ec{equiv[$M.c_1 \sim N.c_2$ : $P \Rightarrow Q$]} - When $P$ is true relating to some memory $m$ and $M.c_1$ and $N.c_2$ terminate in $m$, then after running them in two separate copies of $m$, $Q$ always evaluates to true in the corresponding memories.
\end{description}

In cryptography, security is usually defined by requiring certain properties to hold for all adversaries.
To capture the {\em for all} quantifier, in EasyCrypt, adversaries are defined with abstract procedures, which means the adversaries can do anything without any prescribed strategies.
Working with abstract procedures may require to assume their termination.
In Easycrypt, the idea of termination is modelled by the keyword \ec{islossless}.
We can declare a procedure to be {\em lossless} using the following syntax:
\reducespacetiny
\begin{center}
  \ec{islossless\ $M.c$}
\end{center}
\reducespacetiny
The statement is defined as a pHL judgement \ec{phoare[$M.c$ :\ $\boolT \Rightarrow \boolT$] = 1\%r}, which means the procedure M.c always returns and terminates with a probability 1 (\texttt{1\%r} means real number 1).

All the above constructions are useful for our formalised proof.
In particular, to prove the correctness we used a HL judgement, to prove perfect hiding we used both pRHL to compare the hiding experiment with an artificial experiment and a pHL to finalise the proof, and to prove computational binding we used a pHL judgement to compare the binding experiment to the discrete logarithm experiment.

%% file: sections/contribution.tex

\section{Formal Verification of Pedersen Commitment Scheme}
\label{sec:contribution}
\reducespacetiny

In this section we show how we modelled the generic commitment scheme and its security properties, the Pedersen commitment scheme and we proved its security properties of correctness, perfect hiding and computational binding.

\subsection{Modelling the Scheme}
\label{sec:commitment_formalise}
\reducespacetiny

The abstract commitment scheme is modelled by the following few lines, prototyping the algorithms introduced in Section~\ref{sec:commitments_description}:
\begin{lstlisting}[language=easycrypt,basicstyle=\scriptsize\ttfamily]
module type CScheme = { (* Abstract commitment scheme *)
  proc gen() : value
  proc commit(h: value, m: message) : commitment * openingkey
  proc verify(h: value, m: message, c: commitment, d: openingkey) : bool
}.
\end{lstlisting}
The Pedersen commitment protocol runs between a committer $C$, holding a secret message $m \in \intset_q$ to commit to, and a receiver $R$ who agrees on the group $\pbrackets{\groupset, q, g}$, where $q$ is the order of $\groupset$ and $g$ is its generator, and is defined as the following:
\begin{description}
  \item[Commitment phase]\
    \begin{itemize}
      \item $R$ samples a value $h \inr \groupset$ and sends $h$ to $C$.
      \item $C$ samples an opening value $d \inr \intset_q$, computes the commitment $c = g^d h^m$, and sends $c$ to $R$.
    \end{itemize}
  \item[Verification phase]\
    \begin{itemize}
      \item $C$ sends the pair $\pair{m}{d}$ to $R$.
      \item $R$ checks whether $g^d h^m$ matches to the previously received commitment $c$, and either {\em accepts} if they match or {\em reject} if the do not.
    \end{itemize}
\end{description}

We model the protocol in EasyCrypt as the following:
\begin{lstlisting}[language=easycrypt,basicstyle=\scriptsize\ttfamily]
module Ped : CScheme = { (* Implements a CScheme *)
  proc gen() : value = {
    var x, h;
    x =$ FDistr.dt; (* This randomly samples an element in the field Z_q *)
    h = g^x; (* g is globally defined from the cyclic group theory *)
    return h;
  }

  proc commit(h: value, m: message) : commitment * openingkey = {
    var c, d;
    d =$ FDistr.dt;
    c = (g^d) * (h^m);
    return (c, d);
  }

  proc verify(h: value, m: message, c: commitment, d: openingkey) : bool = {
    var c';
    c' = (g^d) * (h^m);
    return (c = c');
  }
}.
\end{lstlisting}

\subsection{Formalising security properties}
\label{sec:commitment_props_formalise}
\reducespacetiny

The security properties we want to prove are correctness, hiding and binding.
The properties are captured by experiments, which are formally defined below and modelled as in Fig.~\ref{fig:corr_hexp_bexp_easycrypt}.
\begin{definition}[Correctness]
\label{def:com_correctness}
A commitment protocol $\pi$ defined by the triplet $\pbrackets{\alg{G}, \alg{C}, \alg{V}}$ is {\em correct} if for all messages $m \in M$ to commit, let $h = \alg{G}\of{1^n}$ and $(c, d) = \alg{C}\of{h, m}$, then $\alg{V}\of{h, m, c, d} = 1$.
\end{definition}
Loosely speaking, the verification done by the algorithm $\alg{V}$ of any message $m$ committed using the algorithm $\alg{C}$ will always succeed.

For hiding and binding, we have two different adversaries: (i) the {\em unhider} $\alg{U}$, which plays the hiding experiment and has two abstract procedures, one to choose a pair of messages, and another to guess which of the two messages corresponds to a given commitment; (ii) the {\em binder} $\alg{B}$, which plays the binding experiment and has only a procedure to output two different pairs (message, opening value) that bind to the same commitment.
\begin{definition}[Hiding]
\label{def:com_hiding}
Let $\pi = \pbrackets{\alg{G}, \alg{C}, \alg{V}}$ be a commitment protocol.
Then we can define the hiding properties for each polynomial time adversary $\alg{U}$.
\reducespacesmall
\begin{equation*}
  \begin{split}
    \mbox{\em (perfect hiding)      }\quad & \prob{\mathrm{HExp}_{\alg{U}, \pi}\of{n} = 1} = \cfrac{1}{2} \\
    \mbox{\em (computational hiding)}\quad & \exists \negligible.\; \prob{\mathrm{HExp}_{\alg{U}, \pi}\of{n} = 1} \leq \cfrac{1}{2} + \negligible \\
  \end{split}
\end{equation*}
where $\negligible$ is a negligible function.
\end{definition}

\textbf{Hiding experiment}. The hiding experiment $\mathrm{HExp}_{\alg{U}, \pi}$ runs as follows:
\begin{itemize}
 \item The adversary is given the output of $\alg{G}$ and asked to {\em choose} two messages,
 \item the experiment randomly selects one of them and calls $\alg{C}$ to compute its commitment,
 \item the adversary is asked to {\em guess} which one of the two messages the commitment corresponds to, and finally
 \item the experiments outputs $1$ if the guess of the adversary is correct.
\end{itemize}
A commitment protocol satisfies the hiding security property if no adversary exist such that the probability of winning the hiding experiment is (significantly) better than a blind guess.
If this is true, the committer is guaranteed that no information can be inferred by the commitment itself.

\begin{definition}[Binding]
\label{def:com_binding}
Let $\pi = \pbrackets{\alg{G}, \alg{C}, \alg{V}}$ be a commitment protocol.
Then we can define the binding properties for each polynomial time adversary $\alg{B}$.
\reducespacesmaller
\begin{equation*}
  \begin{split}
    \mbox{\em (perfect binding)      }\quad & \exists\, \negligible.\; \prob{\mathrm{BExp}_{\alg{B}, \pi}\of{n} = 1} = 0 \\
    \mbox{\em (computational binding)}\quad & \exists\, \negligible.\; \prob{\mathrm{BExp}_{\alg{B}, \pi}\of{n} = 1} \leq \negligible \\
  \end{split}
\end{equation*}
where $\negligible$ is a negligible function.
\end{definition}

\textbf{Binding experiment}. The binding experiment $\mathrm{BExp}_{\alg{B}, \pi}$ runs as follows:
\begin{itemize}
  \item The adversary is given the output of $\alg{G}$ and asked to {\em bind} two messages to the same commitment value, then
  \item the experiment outputs $1$ if the two messages differ and the commitment is valid for both the messages, that is if both can be verified by calling $\alg{V}$.
\end{itemize}
A commitment protocol satisfies the binding security property if no adversary exist such that the probability of winning the binding experiment is higher than negligible.
If this is true, the receiver is guaranteed that the value committed cannot be changed.

\reducespace
\begin{figure}[htbp]
\caption{Commitment scheme properties. Correctness (top), hiding experiment (left) and binding experiment (right) modelled in EasyCrypt.}
\label{fig:corr_hexp_bexp_easycrypt}
\centering\begin{minipage}{0.6\textwidth}
\begin{lstlisting}[language=easycrypt,basicstyle=\scriptsize\ttfamily]
module Corr (S:CScheme) = {
  proc main(m: message) : bool = {
    var h, c, d, b;

    h = S.gen();
    (c, d) = S.commit(h, m);
    b = S.verify(h, m, c, d);

    return b;
  }
}.
\end{lstlisting}
\end{minipage}
\begin{minipage}[t]{0.5\textwidth}
  \begin{lstlisting}[language=easycrypt,basicstyle=\scriptsize\ttfamily]
module HExp(S:CScheme,U:Unhider) = {
  proc main() : bool = {
    var b, b', m0, m1, h, c, d;

    h = S.gen();
    (m0, m1) = U.choose(h);
    b =$ {0,1};
    (c, d) = S.commit(h, b?m1:m0);
    b' = U.guess(c);

    return (b = b');
  }
}.
  \end{lstlisting}
%
\end{minipage}%
\begin{minipage}[t]{0.5\textwidth}
  \begin{lstlisting}[language=easycrypt,basicstyle=\scriptsize\ttfamily,frame=]
module BExp(S:CScheme,B:Binder) = {
  proc main() : bool = {
    var h, c, m, m', d, d', v, v';

    h = S.gen();
    (c, m, d, m', d') = B.bind(h);
    v  = S.verify(h, m , c, d );
    v' = S.verify(h, m', c, d');

    return v /\ v' /\ (m <> m');
  }
}.
  \end{lstlisting}
%
\end{minipage}
\end{figure}
\reducespace

\subsection{Proofs}
\label{sec:pedersen_formalise}
\reducespacetiny

Relating to the properties modelled in Section~\ref{sec:commitment_formalise}, the Pedersen commitment scheme security properties we have to prove are correctness, perfect hiding, and computational binding.
These security properties rely on the existence of a group $\pbrackets{\groupset, q, g}$ in which the discrete logarithm is {\em hard} to compute (discrete logarithm assumption).

\textbf{Correctness}.
Correctness in EasyCrypt is formalised with a HL judgement: \ec{hoare[ Corr(Ped).main :\ $\boolT \Rightarrow$ res].}
Its proof is straightforward.
The first step is to unfolding the definition of \ec{Corr(Ped).main}, which is the correctness algorithm described in Fig.~\ref{fig:corr_hexp_bexp_easycrypt} instantiated with \ec{Ped} illustrated in Section~\ref{sec:commitment_formalise}.
Then we have $c = g^d h^m$ and $c^{\prime} = g^d h^m$ which are always equal.


\textbf{Perfect hiding}.
In the Pedersen protocol we prove the perfect hiding:
\reducespacesmaller
\begin{equation}
\label{eq:perfect_hiding}
  \forall\, \alg{U}.\quad \prob{\mathrm{HExp}_{\alg{U}, \mathrm{Ped}}\of{\groupset, q, g} = 1} = \cfrac{1}{2}
\end{equation}
In EasyCrypt, we modelled it with the following lemma:
\begin{lstlisting}[language=easycrypt-math]
lemma perfect_hiding: forall (U <: Unhider) &m,
  islossless U.choose $\Rightarrow$ islossless U.guess $\Rightarrow$
  Pr[HExp(Ped, U).main() @ &m : res] = 1%r / 2%r.
\end{lstlisting}
Where \ec{U <: Unhider} is the adversary $\alg{U}$ with abstract procedures \ec{choose} and \ec{guess}, of which we needed to assume they terminate \ec{islossless U.choose} and \ec{islossless U.guess}.

Perfect hiding can be proved by comparing the hiding experiment to an intermediate experiment in which the commitment is replaced by $g^d$ which contains no information about $m_b$.
The experiment is described in Fig.~\ref{fig:fake_hiding}.
\reducespace
\begin{figure}[htbp]
\caption{The intermediate hiding experiment is almost equal to the hiding experiment, but the commitment is replaced by a random group element.}
\label{fig:fake_hiding}
  \SetNoFillComment
  \SetAlgoVlined
\begin{minipage}[t]{0.55\textwidth}
  \begin{algorithm}[H]
    \SetKwBlock{hart}{HInterm\textsubscript{ $\alg{U}$, Ped}}{}

    \hart($\pbrackets{\groupset, q, g}$){
      $h \inr \groupset$; \\
      $\pair{m_0}{m_1} \gets \alg{U}.choose\of{h}$; \\
      $b \inr \zoset$; \\
      $d \inr \intset_q$; \\
      $c \gets g^d$;  \tcp{message independent}
      $b^{\prime} \gets \alg{U}.guess\of{c}$; \\
      return $b = b^{\prime}$;
    }
  \end{algorithm}
\end{minipage}%
\begin{minipage}[t]{0.45\textwidth}
  \begin{lstlisting}[language=easycrypt,basicstyle=\scriptsize\ttfamily]
module HInterm(U:Unhider) = {
  proc main() : bool = {
    var b, b', x, h, c, d, m0, m1;

    x =$ FDistr.dt;
    h = g^x;
    (m0, m1) = U.choose(h);
    b =$ {0,1};
    d =$ FDistr.dt;
    c = g^d; (* message independent *)
    b' = U.guess(c);

    return (b = b');
  }
}.
  \end{lstlisting}
\end{minipage}%
\end{figure}
\reducespacebig

We prove it by first showing that for all adversaries, the probability of winning the hiding experiment is exactly the same as winning the intermediate experiment.
\reducespacetiny
\begin{equation*}
  \forall\, \alg{U}.\quad \prob{\mathrm{HExp}_{\alg{U}, \mathrm{Ped}}\of{\groupset, q, g} = 1} = \prob{\mathrm{HInterm}_{\alg{U}, \mathrm{Ped}}\of{\groupset, q, g} = 1}
\end{equation*}
\reducespacetiny
In code,
\begin{lstlisting}[language=easycrypt-math]
lemma phi_hinterm (U<:Unhider) &m:
  Pr[HExp(Ped,U).main() @ &m : res] = Pr[HInterm(U).main() @ &m : res].
\end{lstlisting}
To prove that, we unfold the two experiments in a pRHL judgement.
The first experiment is automatically instantiated by EasyCrypt as follows:

\SetNoFillComment
\SetAlgoVlined
\begin{algorithm}[H]
  \SetKwBlock{hexp}{HExp\textsubscript{ $\alg{U}$, Ped}}{}

  \hexp($\pbrackets{\groupset, q, g}$){
    $h \inr \groupset$; \\
    $\pair{m_0}{m_1} \gets \alg{U}.choose\of{h}$; \\
    $b \inr \zoset$; \\
    $d \inr \intset_q$; \\
    $c \gets g^d h^{m_b}$; \\
    $b^{\prime} \gets \alg{U}.guess\of{c}$; \\
    return $b = b^{\prime}$;
  }
\end{algorithm}
The proof is done by comparing the execution of the two experiments and is based on the fact that the distribution of $h^{m_b}g^{d}$ is taken over $g^d$.

Then, we proved that for all adversaries, the probability of winning the intermediate experiment is exactly a half.
\reducespacesmaller
\begin{equation*}
  \forall\, \alg{U}.\quad \prob{\mathrm{HInterm}_{\alg{U}, \mathrm{Ped}}\of{\groupset, q, g} = 1} = \frac{1}{2}
\end{equation*}
In EasyCrypt, we have:
\begin{lstlisting}[language=easycrypt-math]
lemma hinterm_half (U<:Unhider) &m:
  islossless U.choose $\Rightarrow$ islossless U.guess $\Rightarrow$
  Pr[HInterm(U).main() @ &m : res] = 1%r/2%r.
\end{lstlisting}
Combining the two lemmas, by transitivity, we prove perfect hiding for Pedersen commitment protocol as in equation \eqref{eq:perfect_hiding}.

\textbf{Computational binding}.
For the Pedersen protocol, we prove the computational binding property.
\begin{equation}
\label{eq:computational_binding}
  \forall\, \alg{B}.\; \exists\, \negligible.\quad \prob{\mathrm{BExp}_{\alg{B}, \mathrm{Ped}}\of{\groupset, q, g} = 1} \leq \negligible
\end{equation}
where $\negligible$ is a negligible function.
The proof is done by a reduction to the discrete logarithm assumption.
In cryptography, proof by reduction usually means to show how to transform an efficient adversary that is able to {\em break} the construction into an algorithm that efficiently solves a problem that is assumed to be hard.
In this proof, the problem assumed to be hard is the discrete logarithm problem~\cite[p.~320]{katz2014introduction}.
We show that if an adversary can break the binding property, then it can output $\pair{m}{d}$ and $\pair{m^{\prime}}{d^{\prime}}$ such that $g^d h^m = g^{d^{\prime}} h^{m^{\prime}}$.
If this is true then the discrete logarithm of $h = g^x$ can be computed by
\begin{equation*}
  x = \frac{d - d^{\prime}}{m^{\prime} - m}.
\end{equation*}

\reducespace
\begin{figure}[htbp]
\caption{The discrete logarithm experiment (left) and an adversary reducing the binding experiment with the Pedersen protocol to the discrete logarithm experiment (right), where \ec{DLogAttacker(B).guess} models $\alg{A}\of{\alg{B}}$.guess.}
\label{fig:dlog_experiment}
\SetNoFillComment
\SetAlgoVlined
\begin{minipage}[t]{0.4\textwidth}
  \begin{algorithm}[H]
    \SetKwBlock{dlexp}{DLog\textsubscript{ $\alg{A}$}}{}

    \dlexp($\of{\groupset, q, g}$){
      $x \inr \intset_q$; \\
      $x^{\prime} \gets \alg{A}.guess\of{g^x}$; \\
      \uIf{$x^{\prime} = \bot$}{%
        $b \gets$ false; \\
      }\Else{%
        $b \gets \pbrackets{x^{\prime} = x}$; \\
      }
      return $b$; \\
    }
  \end{algorithm}
\end{minipage}%
\begin{minipage}[t]{0.6\textwidth}
  \begin{algorithm}[H]
    \SetKwBlock{dlbadversary}{$\alg{A}\of{\alg{B}}$}{}

    \dlbadversary(.guess$\of{h}$){
      $\pbrackets{c, m, d, m^{\prime}, d^{\prime}} \gets \alg{B}.bind\of{h}$; \\
      \uIf{$c = g^d h^m = g^{d^{\prime}} h^{m^{\prime}} \land m \neq m^{\prime}$}{%
        $\displaystyle x \gets \frac{d - d^{\prime}}{m^{\prime} - m}$; \\
      }\Else{%
        $x \gets \bot$; \\
      }
      return $x$; \\
    }
  \end{algorithm}
\end{minipage}
\begin{minipage}[t]{0.4\textwidth}
  \begin{lstlisting}[language=easycrypt,basicstyle=\scriptsize\ttfamily]
module DLog(A:Adversary) = {
  proc main () : bool = {
    var x, x', b;

    x =$ FDistr.dt;
    x' = A.guess(g^x);
    if (x' = None)
      b = false;
    else
      b = (x'= Some x);

    return b;
  }
}.
  \end{lstlisting}
\end{minipage}%
\begin{minipage}[t]{0.6\textwidth}
  \begin{lstlisting}[language=easycrypt,basicstyle=\scriptsize\ttfamily]
module DLogAttacker(B:Binder) : Adversary = {
  proc guess(h: group) : F.t option = {

    var x, c, m, m', d, d';
    (c, m, d, m', d') = B.bind(h);
    if ((c = g^d * h^m) /\
        (c = g^d' * h^m') /\ (m <> m'))
      x = Some((d - d') * inv (m' - m));
    else
      x = None;

    return x;
  }
}.
  \end{lstlisting}
\end{minipage}
\end{figure}
\reducespace

We capture the reduction by two modules in EasyCrypt (Fig.~\ref{fig:dlog_experiment}).
A small technical subtlety is that since the adversary is abstractly defined, it can return $m = m^{\prime}$ with some probability. This can cause division by zero.
Therefore, we check the output from the adversary to avoid it.
Formally, the adversary assumed to break the binding experiment is $\alg{B}$ and we construct an adversary $\alg{A}$ to break the discrete logarithm experiment with equal probability of success:
\begin{equation*}
  \begin{split}
  \forall\, \alg{B}.\quad \prob{\mathrm{BExp}_{\alg{B}, \mathrm{Ped}}\of{\groupset, q, g} = 1} & = \prob{\mathrm{DLog}_{\alg{A}(\alg{B})}\of{\groupset, q, g} = 1}
  \end{split}
\end{equation*}
The above is captured in EasyCrypt by the lemma:
\begin{lstlisting}[language=easycrypt-math]
lemma computational_binding: forall (B <: Binder) &m,
  Pr[BExp(Ped, B).main() @ &m : res] =
  Pr[DLog(DLogAttacker(B)).main() @ &m : res].
\end{lstlisting}
To prove the lemma, we unfolded the experiments as much as possible, i.e. up to abstractions, in a pRHL judgement which created an equivalence of the two experiments in the sense illustrated in Section~\ref{sec:easycrypt}.
The binding experiment is automatically unfolded to the following.

\SetNoFillComment
\SetAlgoVlined
\begin{algorithm}[H]
  \SetKwBlock{bexp}{BExp\textsubscript{ $\alg{B}$, Ped}}{}

  \bexp($\pbrackets{\groupset, q, g}$){
    $h \inr \groupset$; \\
    $\pbrackets{c, m, d, m^{\prime}, d^{\prime}} \gets \alg{B}.bind\of{h}$; \\
    $v          \gets c = g^d            h^m           $ \\
    $v^{\prime} \gets c = g^{d^{\prime}} h^{m^{\prime}}$ \\
    return $v \land v^{\prime} \land m \neq m^{\prime}$;
  }
\end{algorithm}

The automatic tactics could not automatically prove the lemma, as the expression $\pbrackets{d - d^{\prime}}/\pbrackets{m^{\prime} - m}$ used by the attacker $\alg{A}$ (modelled as \ec{DLogAttacker}) in the DLog experiment was too complex to be automatically used by the prover into the binding experiments and needed to be manually guided.

Assuming that the discrete logarithm is hard, then the probability of the experiment $\mathrm{BExp}_{\alg{B}, \mathrm{Ped}}\of{\groupset, q, g}$ returning $1$ must be negligible.
Finally,
\begin{equation*}
  \forall\, \alg{B}.\; \exists\, \negligible.\quad \prob{\mathrm{BExp}_{\alg{B}, \mathrm{Ped}}\of{\groupset, q, g} = 1} \leq \negligible
\end{equation*}
which is the definition of computational binding we gave in equation \eqref{eq:computational_binding}.

%% file: sections/conclusion.tex

\section{Conclusion and future work}
\label{sec:conclusion}
\reducespacetiny

In this paper, we showed how EasyCrypt can be used for formally verifying practical cryptographic primitives and automatising mechanised proofs. With a game based approach, we could construct fully mechanised proofs of the security properties of the Pedersen commitment protocol, a building block primitive for many cryptographic protocols.

Composability is a desirable property of cryptographic protocols. When designing a protocol, we often want to guarantee that the composition of the protocol does not break the required security properties. In cryptography, advanced theories like Universal Composability have been proposed. As a future work, we will investigate how to enable machine-aided proofs for composability.

